\documentclass{aa}
\usepackage{graphicx}
\usepackage{verbatim}
\usepackage{appendix}
\usepackage{epsfig}
\usepackage{ulem}
\usepackage{color}
\usepackage{amsmath}    
\usepackage{mathtools}
\usepackage{threeparttable}
\usepackage{booktabs}
\usepackage{txfonts}
\usepackage[dvipsnames]{xcolor}
\usepackage[pdfstartview=FitH,
CJKbookmarks=true,
bookmarksnumbered=true,
bookmarksopen=true,
colorlinks,
pdfborder=001,
linkcolor=blue,
anchorcolor=blue,
citecolor=blue,
]{hyperref}
\hypersetup{hidelinks}

\begin{document} 

\newcommand{\andres}[1]{\textcolor{ForestGreen}{#1}}
\newcommand{\tobias}[1]{\textcolor{blue}{#1}}
\newcommand{\corr}[1]{\textcolor{red}{#1}}
\newcommand{\elena}[1]{\textcolor{BurntOrange}{#1}}

   \title{Mixing fraction in classical novae}


   \author{Yunlang Guo\inst{1,}\inst{2}  
                        \and
          Chengyuan Wu \inst{1,}\inst{2} 
           \and
          Bo Wang\inst{1,}\inst{2}
          }

   \institute{Yunnan Observatories, Chinese Academy of Sciences, Kunming 650216, China\label{inst1}\\
              \email{wuchengyuan@ynao.ac.cn; wangbo@ynao.ac.cn}
         \and
             University of Chinese Academy of Sciences, Beijing 100049, China\label{inst2} 
             }

   \date{Received xxx, aaaa; accepted yyy, aaaa}

 
  \abstract
   {Classical novae are powered by thermonuclear runaways occurring on the surface of accreting white dwarfs (WDs).
        In the observations, the enrichments of heavy elements in nova ejecta have been detected,
        indicating a mixing process between the accreted matter and
        the matter from the outer layers of the
        underlying WDs prior to nova outbursts. However, the mixing fraction in classical novae is still uncertain.}
        {The purpose of this article is to investigate some elemental abundance ratios during nova outbursts that can be used to estimate the WD mixing fraction in classical novae.}         
   {By considering different WD mixing fractions with the stellar evolution code Modules for Experiments in Stellar Astrophysics,
        we carried out a series of simulations of nova outbursts, in which the initial CO WD masses range from $0.7-1.0\,M_\odot$.}
        {We identified four elemental abundance ratios
                (i.e. $\rm (H+He)/\sum CNO$, $\rm (H+He)/Ne$, $\rm \sum CNO/Mg,$ and $\rm \sum CNO/Si$)
                that satisfy the conditions for determining the WD mixing fraction,
                in which $\rm (H+He)/\sum CNO$ is the most suitable mixing meter.
        We also estimated the WD mixing fraction in some representative classical novae.
        Additionally, we found that a higher metallicity (i.e. higher WD mixing fraction) is preferentially accompanied by a longer $t_{\rm 2}$
        (the time of decline by two magnitudes from peak luminosity) during nova outbursts.
        Our results can be used to constrain the mixing process in classical novae.
}
    {} 


   \maketitle
%
\section{Introduction}\label{introduction}

It is generally thought that classical novae are powered by thermonuclear runaways
occurring on the surface of white dwarfs (WDs) in close binaries,
in which the WDs accrete matter from a main-sequence
companion through
the inner Lagrangian point when the companion fills
its Roche lobe
(e.g. Jos{\'e} 2016; Chomiuk et al. 2021).
Classical novae are suggested as noteworthy contributors to the Galactic chemical evolution,
providing about $0.3\%$ of the interstellar matter in the Galaxy.
The nuclides $^{13}\rm C$, $^{15}\rm N,$ and $^{17}\rm O$ in the interstellar medium are thought to be mainly synthesised by classical novae (e.g. Jos{\'e} \& Hernanz 1998; Denissenkov et al. 2014).
In addition, classical novae are suggested to be the progenitors of type Ia supernovae
and accretion-induced collapse events (see, e.g. Wang 2018; Wang \& Liu 2020).

Observationally, the enrichments of intermediate-mass elements
(e.g. C, N, O, Ne, Na, Mg, and Al)
have been detected in nova ejecta,
indicating that the accreted shell is mixed with the outer layers of the underlying WD
(e.g. Livio \& Truran 1994; Gehrz et al. 1998).
Two types of WDs in nova systems can be identified by the spectroscopic studies for nova ejecta: carbon-oxygen (CO) WDs, and oxygen-neon (ONe) WDs.
Some mechanisms from one-dimensional models have been proposed to explain the mixing process,
including diffusion-induced convection (e.g. Prialnik \& Kovetz 1984; Fujimoto \& Iben 1992),
shear mixing (e.g. Durisen 1977; Kippenhahn \& Thomas 1978; MacDonald 1983; Kutter \& Sparks 1987; Fujimoto 1988),
and convective overshoot-induced flame propagation (see, e.g. Woosley 1986).
For multidimensional models, the mechanisms include mixing by resonant gravity waves (e.g. Rosner et al. 2001; Alexakis et al. 2004)
and Kelvin-Helmholtz instabilities (e.g. Glasner \& Livne 1995; Glasner et al. 1997, 2005, 2007, 2012; Casanova et al. 2010, 2011, 2016, 2018).

The chemical abundances in nova ejecta have been proposed as a way to reveal the characteristics of the nova outbursts.
Jos{\'e} \& Hernanz (1998) studied five classical novae by simulating 14 evolutionary sequences of CO/ONe nova models,
spanning a range of WD masses and WD mixing fractions.
Their results indicate that the significant $^{19}\rm F$, $^{35}\rm Cl$,
and low O/N, C/N ratios can be reproduced in the nova systems with a massive ONe WD.
Downen et al. (2013) found some useful elemental abundance ratios strongly correlated with the peak temperatures during nova outbursts (e.g. $\rm N/O$, $\rm N/Al,$ and $\rm O/S$),
providing a way to determine the WD mass.
Kelly et al. (2013) identified some elemental abundance ratios
(e.g. $\rm \sum CNO/H$, $\rm Ne/H$, $\rm Mg/H$, $\rm Al/H,$ and $\rm Si/H$)
to estimate the mixing fraction for ONe novae and
provided some constraints on the study of the WD mixing process.
These ratios show a strong dependence on the mixing fraction,
but
they are not significantly affected by the WD mass.

The enrichments of He have also been detected in classical novae
(e.g. GQ Mus, HR Del, and RR Pic),
indicating that there could be a mixing process between the He shell and the accreted material
(e.g. Truran \& Livio 1986; Gehrz et al. 1998).
Starrfield et al. (1998) suggested that the accreted material may have mixed with the He layer
by diffusion or accretion-driven shear mixing before the dredge-up of the WD
material.
In addition, it is also possible that the WD material has enriched in the He layer due to the previous evolutionary history,
which means that the mixing of the accreted material and the He layer can naturally increase the abundance of He and WD matter.
By assuming the mixing of the He shell and the accreted material in multidimensional simulations,
Glasner et al. (2012) estimated that the He mixing fraction is about $20\%$.
Mixing with different chemical substrates has also been reported in Casanova et al. (2016, 2018).
Guo et al. (2021b) recently reproduced the chemical abundances in some novae with He enrichments by considering the
He and WD mixing process.
They found that the He mixing level in classical novae is lower than $30\%$.

The purpose of this article is to study several elemental abundance ratios
that can be used to estimate the WD mixing fraction in CO novae.
These ratios should be sensitive to the WD mixing degree
but not affected by the WD mass.\footnote{
Although the WD mixing fraction would be affected by the WD mass (e.g. Casanova et al. 2018),
we focus on the measurement of the WD mixing degree,
thus the mixing meters must only be sensitive to the mixing degree.
The accuracy of the measurement is reduced if the WD mass has an effect on the mixing meters.}
In addition, it is worth noting that these ratios cannot be affected by the He mixing that could occur in CO novae because the abundances of H and He in nova ejecta
        are sensitive to the
He mixing fractions (see Guo et al. 2021b).
This paper is organised as follows.
In Section 2 we introduce the basic assumptions and methods for numerical calculations.
In Section 3 we identify some WD mixing meters
and estimate the WD mixing fraction in several representative CO novae.
In addition, we show the relation between $t_{\rm 2}$
(the time of decline by two magnitudes from peak luminosity)
and the metallicity in nova ejecta.
Finally, we present a discussion in Section 4 and a summary in Section 5.

\section{Numerical methods and assumptions}\label{methods}
\subsection{Nova models}
The stellar evolution code we used in our simulations is called Modules for Experiments in Stellar Astrophysics
(MESA, version 10398; see Paxton et al. 2011, 2013, 2015, 2018).
It is widely accepted that the properties of classical novae (e.g. light curves,
timescales, nucleosynthesis, and ejecta mass) depend on
the WD mass ($M_{\rm WD}$), the mass-accretion rate ($\dot M_{\rm acc}$),
the initial luminosity of WDs, and the composition of accreted matter
(e.g. Yaron et al. 2005; Wolf et al. 2013; Rukeya et al. 2017; Wang 2018).
In the present work,
we created the CO WD models with an initial mass range of $0.7-1.0\,M_\odot$.\footnote{
                The WDs with mass $\lesssim 0.6\, M_\odot$ are considered to be
                hybrid HeCO WDs with a CO-rich core surrounded by a $\sim0.1\, M_\odot$ He-rich mantle
                (e.g. Dan et al. 2012; Yungelson \& Kuranov 2017; Liu et al. 2017).
                According to current stellar evolution theories, the composition of massive WDs is ONe instead of CO
                if the WD mass is higher than $\sim1.05\,M_\odot$ (see Siess 2007).
}
Following Guo et al. (2021b),
our initial WD models have a luminosity of $10^{-2}\,L_\odot$,
and the accretion rate ($\dot M_{\rm acc}$) in the nova models is $2 \times 10^{-9}\,M_\odot \rm yr^{-1}$.
The nuclear reaction network is coupled by 86 isotopes from $^{1}\rm H-^{48}\rm Ca$, including 794 nuclear reactions.
In addition,
we did not consider the influence of rotation and convective overshooting in this work.

We adopted the super-Eddington wind as the mass-loss mechanism during nova outbursts,
in which the super-Eddington luminosity ($L_{\rm Edd}$) and the wind mass-loss rate ($\dot M$) can be expressed as follows
(e.g. Denissenkov et al. 2013; Wang et al. 2015; Wu et al. 2017; Guo et al. 2021a): 
\begin{equation}
L_{\rm Edd} = \frac{4\pi GcM_{\rm WD}}{\tau},
\end{equation}
\begin{equation}
\dot M = \frac {2\eta_{\rm Edd}(L-L_{\rm Edd})}{\upsilon_{\rm esc}^2},
\end{equation}
where $G$, $c,$ and $\tau$ are the gravitational constant, the vacuum speed of light, and the Rosseland mean opacity, and $\eta_{\rm Edd}$ and $\upsilon_{\rm esc}$ are the super-Eddington wind factor and escape velocity, respectively.
The super-Eddington wind is triggered when the WD luminosity exceeds the Eddington luminosity,
and all the radiant energy exceeding Eddington luminosity is used to eject matter (i.e. $\eta_{\rm Edd}=1$). 

\subsection{White dwarf mixing}
We adopted the pre-mixed model described in Politano et al. (1995),
in which the material accreted by WDs is assumed to be a mixture of the companion star material (solar abundances) and the outermost
layers of the underlying CO WDs.
Different WD mixing degrees can be modified by artificially giving the various abundance of the WD material in the mixture.
In our simulations,
we assumed that the WD material mixed with the accreted layer is composed of $X(^{12}\rm C)$ = 0.495, $X(^{16}\rm O)$ = 0.495, and $X(^{22}\rm Ne)$ = 0.01
(e.g. Jos{\'e} \& Hernanz 1998; Guo et al. 2021b).
In order to understand the dependence of chemical abundances in nova ejecta on the WD mixing level,
we carried out a series of calculations for our nova models with different WD mixing fractions ($f_{\rm WD}$) $\leq 40\%$.
\subsection{Helium mixing}
In order to test whether the He mixing has an effect on the WD mixing meters,
we calculated some nova models, including both the He and WD mixing
(see the second paragraph of Sect. 3.1.1 and Fig.\,2 for details).
Following the studies of Guo et al. (2021b),
we mixed the accreted material with a given mass fraction of He layer and WD material.
Similar to the method of the WD mixing,
the He mixing degrees can be modified by
considering the various abundance of the He-rich material in the mixture.
If we only consider the He mixing during the outbursts,
the He mass fraction in the already mixed envelope ($X_{\rm He}$) is contributed by
the mixed He shell and the accreted material.
We assumed the mass ratio of the mixed He shell to the already mixed envelope as the He mixing fraction ($f_{\rm He}$),
and set the He abundance in accreted material and He shell was $28\%$ and $100\%$,
respectively.
Thus, the He mass fraction can be calculated by
$X_{\rm He}=f_{\rm He} + 0.28 \times (1-f_{\rm He})$,
in which $0.28 \times (1-f_{\rm He})$
represents the He abundance contributed by the accreted material.
Similarly, if we further consider both He mixing and WD mixing during the outbursts,
then $X_{\rm He}=[f_{\rm He} + 0.28 \times (1-f_{\rm He})] \times (1-f_{\rm WD})$,
in which $f_{\rm WD}$ is the given WD mixing fraction.\footnote{
                We assumed that the accreted material mixes first with the He shell
                and then with the WD material (e.g. Starrfield et al. 1998).
}
\begin{figure}
        \centering\includegraphics[width=\columnwidth*5/5]{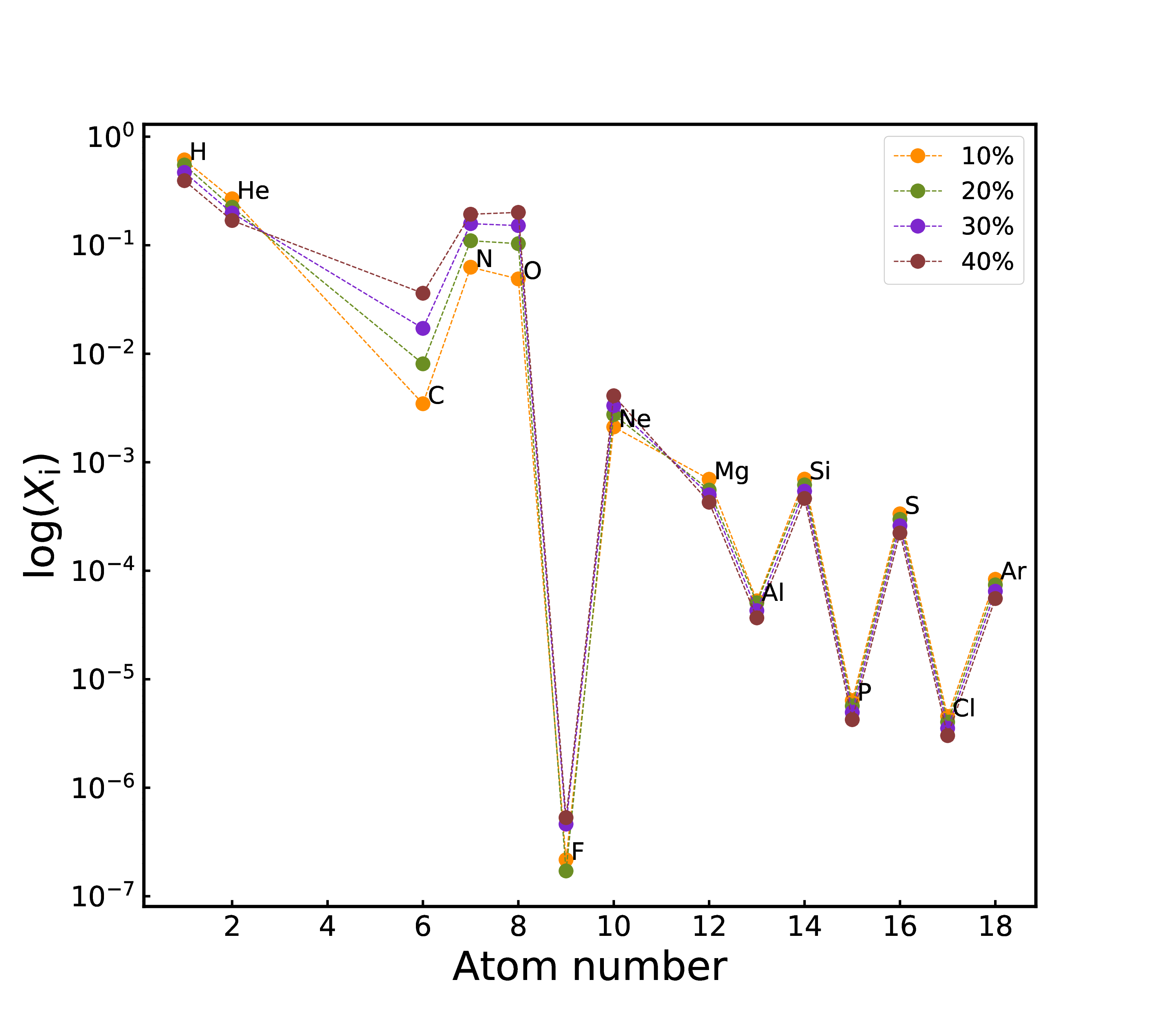}
        \caption{Elemental abundances (mass fraction) in nova ejecta vs. atomic number, in which we set the WD mixing fraction $\leq40\%$ and the WD mass to be $0.9\,M_\odot$.
        Differently coloured lines represent different WD mixing fractions. }
        \label{fig:com_mixing-levels4}
\end{figure}
\section{Numerical results}\label{comparison}
\subsection{WD mixing meters}

We performed a series of simulations for nova models with different WD masses and WD mixing degrees.
Table 1 lists the main properties of initial nova models computed in this work,
as well as the characteristics of the outbursts.
Fig.\,1 presents some chemical abundances (i.e. H-Ar) in nova ejecta that change with different WD mixing fractions,
in which we set the initial WD mass to be $0.9\,M_\odot$.
Obviously, C, N, O, F and Ne exhibit a strong monotonic increase with the WD mixing levels,
while other elements show the opposite trend.
The abundances of some elements are relatively low (e.g. F, P, and Cl) or they are rarely detected in CO nova ejecta (e.g. Al, S, and Ar).
Thus, these elements cannot be used as the WD mixing meters.
We studied the elemental abundance ratios that
satisfy the conditions for determining the WD mixing degree,
for instance $\rm (H+He)/\sum CNO$, $\rm (H+He)/Ne$, $\rm \sum CNO/Mg,$ and $\rm \sum CNO/Si$.

\begin{table}
        \centering
        
        \caption{Initial parameters of nova models and the characteristics of the outbursts:
WD mixing fraction ($f_{\rm WD}$) and WD mass ($M_{\rm WD}$);
                $T_{\rm peak}$ is the peak temperature at the burning shell;
                $M_{\rm ej}$ and  $L_{\rm max}$ are the ejected mass and the peak luminosity during nova outbursts, respectively; and
                $Z$ is the metallicity in nova ejecta.} 
        
        \begin{tabular}{ l  c c c ccc c c c  l }
                \toprule
                \hline \\
                Mixing fraction &      $M_{\rm WD}$  &$T_{\rm peak}$   &$M_{\rm ej}$                     &$L_{\rm max}$ & $Z$\\
                ($f_{\rm WD}$) &   ($M_\odot$) & (GK)                  &$(10^{-5}\,M_\odot)$&$(10^{4}\,L_\odot)$&\\
                \hline 
                &0.7   &0.113                     &4.89                                        &1.82           &0.1204\\
                $10\%$          &0.8                            &0.122                &3.83                                     &2.57           &0.1201\\
                &0.9                            &0.134                    &3.23                                 &2.89           &0.1196\\
                &1.0                            &0.148                    &2.20                                 &3.09           &0.1192\\
                \hline
                &0.7    &0.114                &5.75                                    &2.02           &0.2264\\
                $20\%$  &0.8                            &0.124                    &3.84                                 &2.54           &0.2268\\
                &0.9                            &0.135                    &2.44                                 &3.00           &0.2261\\
                &1.0                            &0.147                    &1.60                                 &3.70           &0.2256\\
                \hline
                &0.7    &0.114         &5.22                                      &2.04         &0.3285\\
                $30\%$          &0.8                            &0.124                    &3.43                                 &2.63           &0.3320\\
                &0.9                            &0.135                    &2.20                                 &3.06           &0.3331\\
                &1.0                &0.145                     &1.41                                   &3.98           &0.3327 \\
                \hline 
                &0.7   &0.114                          &4.66                                   &2.20           &0.4274\\
                $40\%$          &0.8                           &0.124                          &3.10                                   &2.85           &0.4239\\
                &0.9                           &0.136                          &2.14                                   &3.45           &0.4368\\
                &1.0                           &0.144                          &1.28                                   &4.27           &0.4386\\       
                \hline
        \end{tabular}
\end{table}

\subsubsection{$\rm (H+He)/\sum CNO$}
Previous studies suggested that the peak temperature attained in the H-burning shell does not exceed
$4 \times 10^8$K during nova outbursts
(e.g. Livio \& Truran 1994; Denissenkov et al. 2013).
The sum abundance of C, N, and O ($\rm \sum CNO$) would not change significantly in this temperature regime
(e.g. Truran \& Livio 1986; Jos{\'e} \& Hernanz 1998).
For this reason, Kelly et al. (2013) used $\rm H/\sum CNO$ to measure the WD mixing degree in ONe novae.

\begin{figure}
        \centering\includegraphics[width=\columnwidth*5/5]{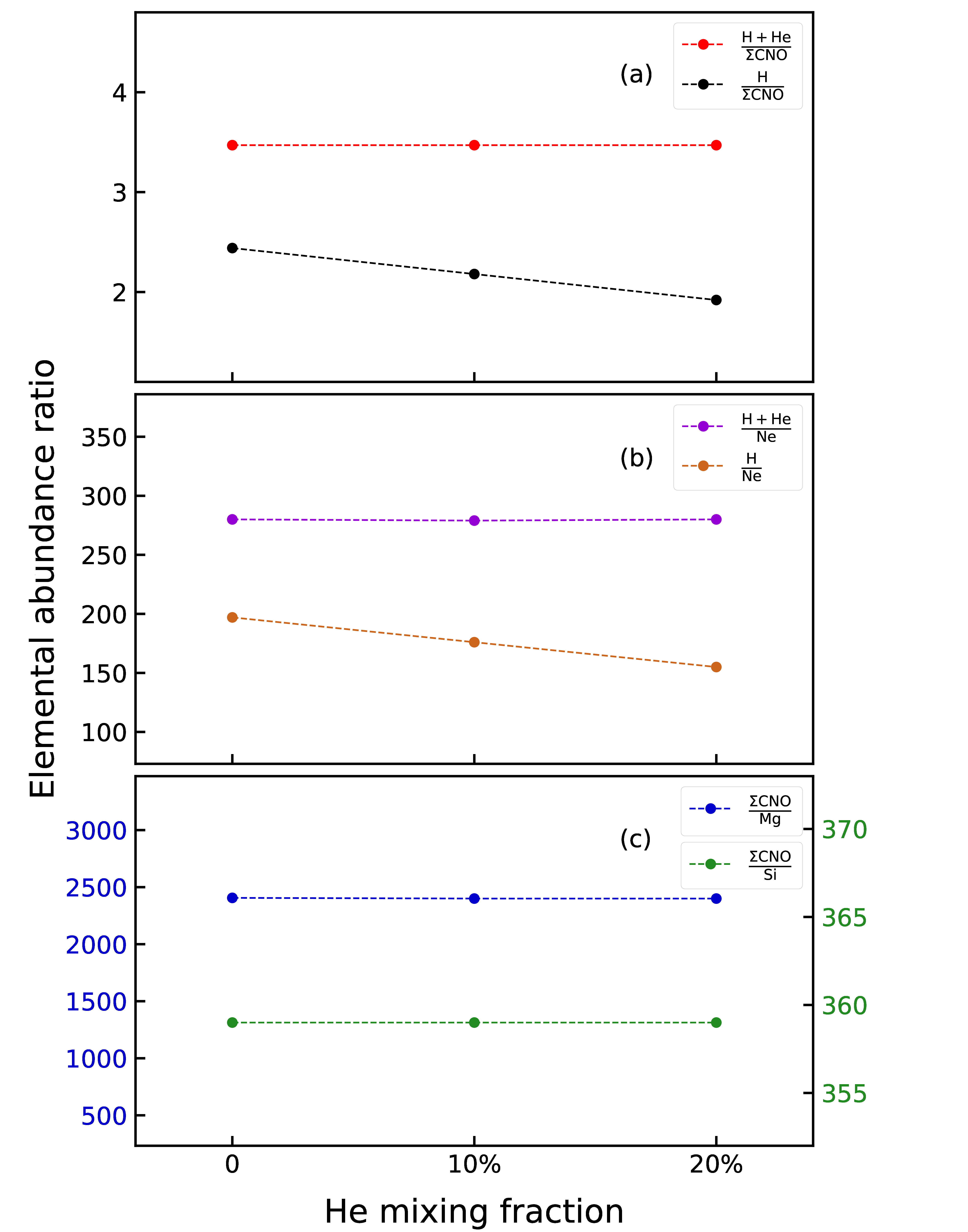}
        \caption{Elemental abundance (mass fraction) ratios that may be candidates for the WD mixing meters vary with different He mixing degrees, in which we set the WD mass to be $0.7\,M_\odot$ and the WD mixing fraction to be $20\%$.
                Panel (a): Results of $\rm (H+He)/\sum CNO$ and $\rm H/\sum CNO$.
                Panel (b): Results of $\rm (H+He)/Ne$ and $\rm H/Ne$.
                Panel (c): Results of $\rm \sum CNO/Mg$ and $\rm \sum CNO/Si$.}
        \label{fig:kkHe-mixing2}
\end{figure}

However, it is not negligible to verify whether the WD mixing meters are affected by the He mixing that may occur in CO novae.
We therefore performed a series of calculations for nova models with a He mixing fraction ranging from $0-20\%$,
where we set the WD mass to be $0.7\,M_\odot$ and the WD mixing fraction to be $20\%$.
Fig.\,2 shows the variation in elemental abundance (mass fraction) ratios with different He mixing degrees.
The value of $\rm H/\sum CNO$ is significantly affected by the He mixing fraction,
while the effect can be eliminated by choosing $\rm (H+He)/\sum CNO$ (see Fig.\,2a). 
This is because for a given WD mixing degree,
the abundance of H+He and heavy elements in nova ejecta is hardly changed by the He mixing process
(see Table 3 in Guo et al. 2021b).
\begin{table*}
        \centering
        
        \caption{Comparison of the neon abundance of Jos{\'e} \& Hernanz (1998) and this work.}
        \begin{tabular}{ l  c c c ccc c c c  l }
                \toprule
                \hline \\
                &       $M_{\rm WD}$                            &&$f_{\rm WD}$&& Ne        &&$Z$  &&$\rm Ne/Z$ \\
                &   ($M_\odot$)                 &&($\%$)          &&     &&                              &&      \\
                \hline 
                &0.80           &&25   &&3.80e-3        &&0.280            &&0.014       \\
                &0.80           &&50   &&5.82e-3        &&0.530            &&0.011       \\                              
                Jos{\'e} \& Hernanz (1998)              &1.00           &&50   &&5.85e-3     &&0.530            &&0.011      \\                              
                &1.15           &&25   &&3.60e-3    &&0.280                &&0.013       \\              
                &1.15           &&50   &&5.77e-3        &&0.540            &&0.011  \\
                &1.15           &&75   &&7.82e-3        &&0.790            &&0.010       \\
                \hline
                &0.70           &&20   &&2.75e-3        &&0.227            &&0.012       \\
                &0.90           &&10   &&2.11e-3        &&0.120            &&0.018       \\
                This work                                       &0.90           &&20   &&2.36e-3     &&0.226            &&0.010      \\
                &0.90           &&30   &&3.14e-3        &&0.333        &&0.010  \\
                &0.90           &&40   &&4.10e-3        &&0.437            &&0.010       \\
                &1.00           &&20   &&2.39e-3        &&0.226            &&0.011  \\
                
                \hline 
        \end{tabular}
\end{table*}

Fig.\,3 represents the changes in four elemental abundance ratios with different WD mixing fractions and peak temperatures,
where the peak temperature corresponds to the WD mass in the range of $0.7-1.0\,M_\odot$.
As shown in Fig.\,3a,
$\rm (H+He)/\sum CNO$ is weakly affected by the peak temperature but strongly depends on the WD mixing fraction,
indicating that $\rm (H+He)/\sum CNO$ can be used to determine the WD mixing fraction.
\begin{figure}
        \centering\includegraphics[width=\columnwidth*10/10]{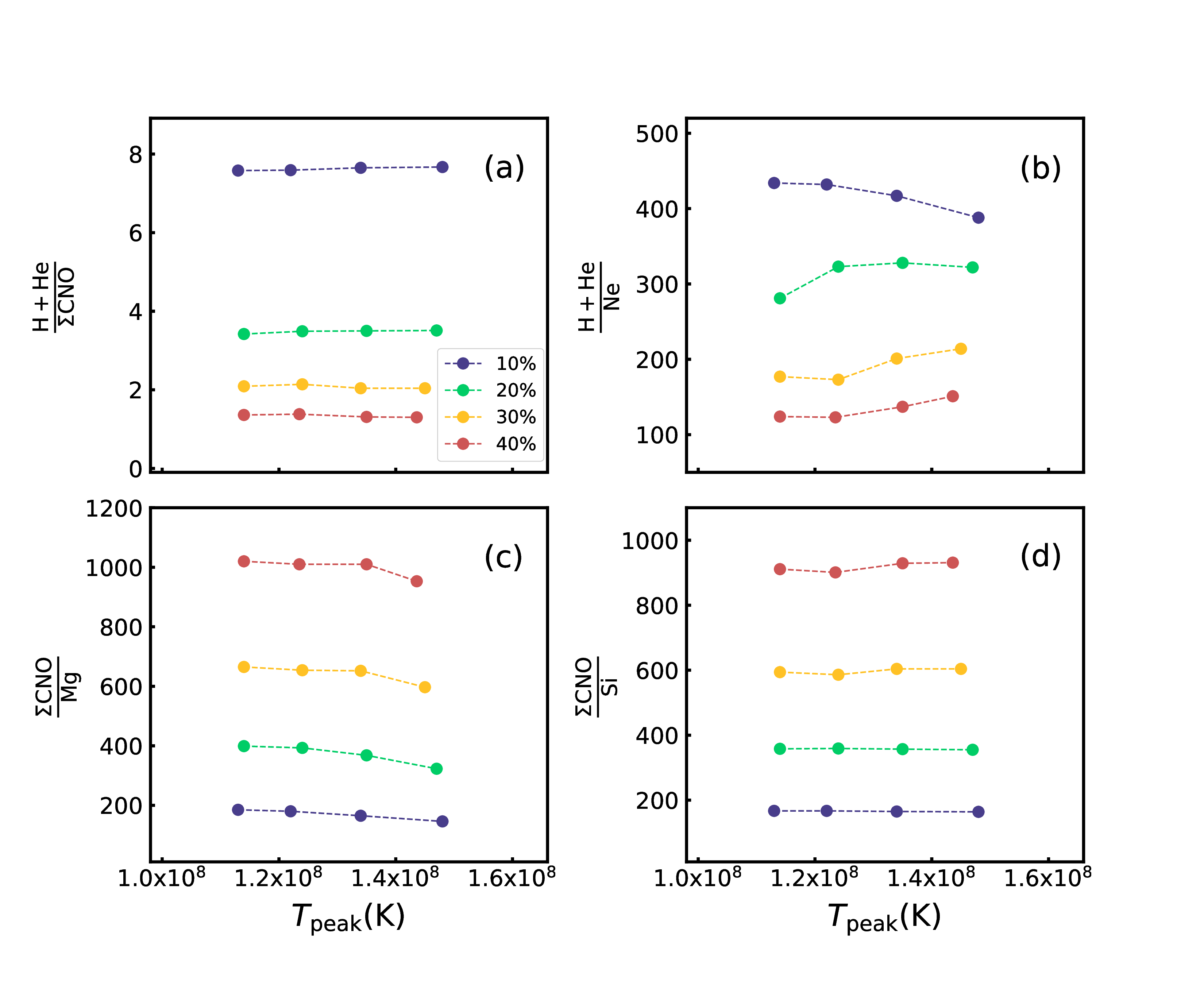}
        \caption{Peak temperature vs. the elemental abundance ratios that can be used to determine the WD mixing fraction,
                where the peak temperature corresponds to the WD mass in the range of $0.7-1.0\,M_\odot$.
                Different colours represent different WD mixing fractions.
                Panel (a): Results of $\rm (H+He)/\sum CNO$.
                Panel (b): Results of $\rm (H+He)/Ne$.
                Panel (c): Results of $\rm \sum CNO/Mg$.
                Panel (d): Results of $\rm \sum CNO/Si$.}
        \label{fig:ding3}
\end{figure}

\begin{figure*}
        \centering\includegraphics[width=\columnwidth*16/10]{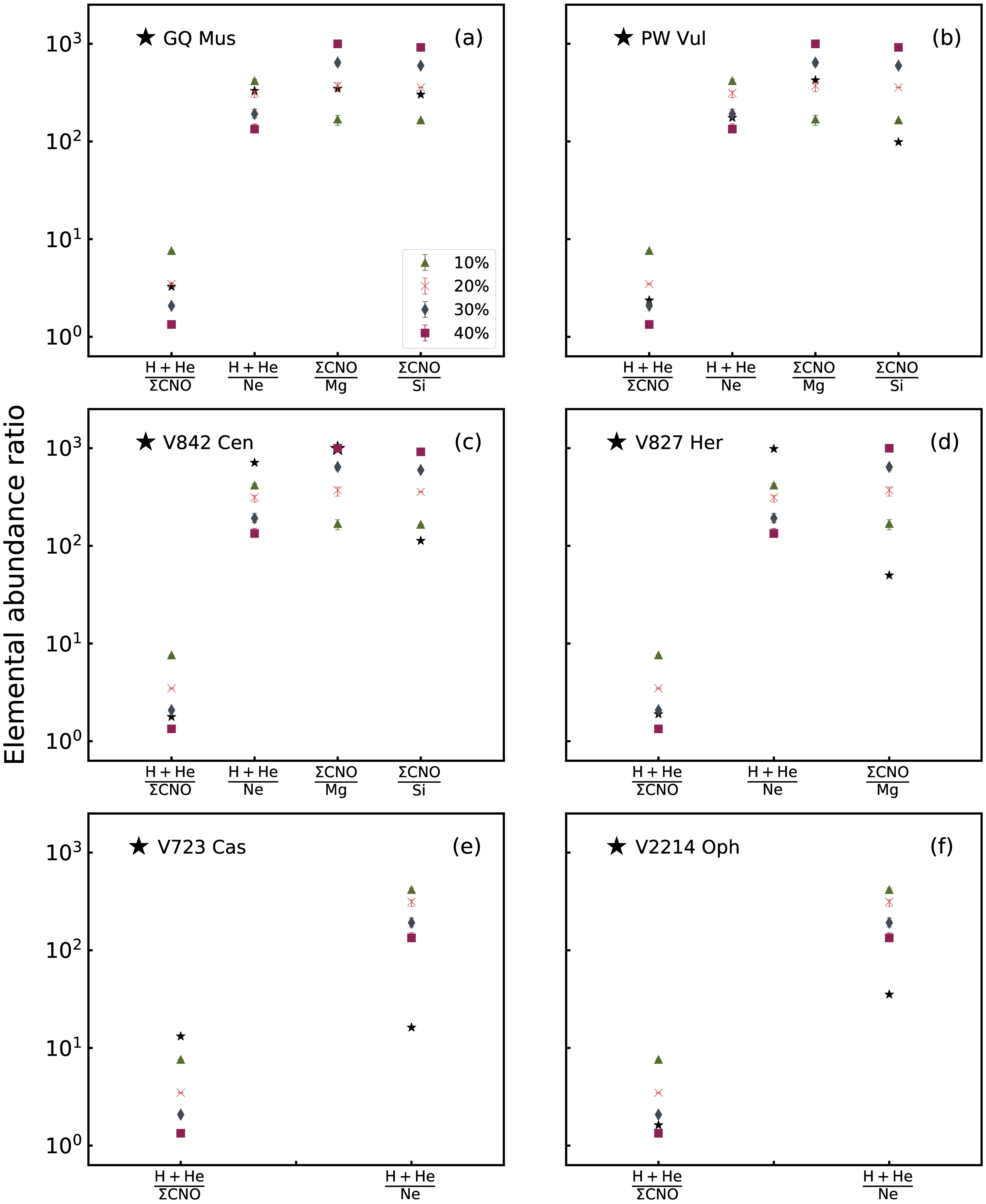}
        \caption{WD mixing fractions in some representative classical novae determined by different elemental abundance ratios.
                Different marks represent the WD mixing degree from $10\%-40\%$.
                The black stars refer to the observed data.
                The error bar represents the effect of the peak temperature on the elemental abundance ratios.
                Panel (a): Results of GQ Mus (Morisset \& Pequignot 1996).
                Panel (b): Results of PW Vul (Andre{\"a} et al. 1994).
                Panel (c): Results of V842 Cen (Andre{\"a} et al. 1994).
                Panel (d): Results of V827 Her (Andre{\"a} et al. 1994).
                Panel (e): Results of V723 Cas (Iijima 2006).
                Panel (f): Results of V2214 Oph (Andre{\"a} et al. 1994).}
        \label{fig:heerror3}
\end{figure*}
\subsubsection{$\rm (H+He)/Ne$}
Table 2 lists the Ne abundance in different nova models,
including the results of Jos{\'e} \& Hernanz (1998) and this work.
These simulations show that the abundance of Ne hardly changes with the WD mass for a given WD mixing fraction.
As shown in Fig.\,2b, $\rm (H+He)/Ne$ is more suitable for the WD mixing meter than $\rm H/Ne$
because of the weak dependence of the former on the He mixing fraction.
Meanwhile, $\rm (H+He)/Ne$ shows a weak dependence on the peak temperature,
whereas sensitive to the WD mixing fraction (see Fig.\,3b).
Thus, $\rm (H+He)/Ne$ can also be a WD mixing meter.

\subsubsection{$\rm \sum CNO/Mg$ and $\rm \sum CNO/Si$}
Downen et al. (2013) studied the ratio of final to initial elemental abundances during nova outbursts that change with different WD masses.
Their results show that the abundances of Mg and Si are hardly affected by the WD mass.
Fig.\,2c shows that $\rm \sum CNO/Mg$ and $\rm \sum CNO/Si$ are not affected by the He mixing fraction.
We present the changes in $\rm \sum CNO/Mg$ and $\rm \sum CNO/Si$ with peak temperature in Fig.\,3c-d, respectively.
These two abundance ratios are clearly weakly dependent on the WD mass,
but strongly depend on the WD mixing degree.
This means that $\rm \sum CNO/Mg$ and $\rm \sum CNO/Si$
meet the conditions for determining the WD mixing degree.
\subsection{Comparison to observations}
Fig.\,4 shows the values of these four ratios with different WD mixing degrees,
as well as the estimation of WD mixing levels in six representative novae,
that is, GQ Mus, PW Vul, V842 Cen, V827 Her, V723 Cas, and V2214 Oph.
GQ Mus is a classical nova,
showing signs of the He and metal enrichments.
According to the studies of Guo et al. (2021b),
the WD and He mixing level in this nova are $20\%$ and $30\%$, respectively.
Similarly, the simulations in this work show that the WD mixing fraction in GQ Mus is about $20\%$.
For nova PW Vul, $\rm (H+He)/\sum CNO$ and $\rm (H+He)/Ne$ imply that the WD mixing fraction is about $30\%$,
which is consistent with the results of Jos{\'e} \& Hernanz (1998).

In the case of nova V842 Cen,
the WD mixing fraction in the range of $30\%-40\%$ is estimated by $\rm (H+He)/\sum CNO$,
while $\rm (H+He)/Ne$ implies a value lower than $10\%$.
For nova V827 Her, $\rm (H+He)/\sum CNO$ indicates a WD mixing fraction of $30\%$,
while $\rm (H+He)/Ne$ shows a value lower than $10\%$.
Nova V723 Cas shows a WD mixing fraction below $10\%$ by the estimation of $\rm (H+He)/\sum CNO$,
while a value far greater than $40\%$ is suggested by $\rm (H+He)/Ne$.
V2214 Oph is also a classical nova showing He enrichments,
and the He mixing fraction is about $13\%$ (see Guo et al. 2021b).
For this nova, $\rm (H+He)/\sum CNO$ implies a WD mixing degree in the range of $30\%-40\%$,
while $\rm (H+He)/Ne$ indicates that the value is higher than $40\%$.
In addition,
DQ Her exhibits a high enrichment of CNO,
but only H, He, C, N, and O were determined by Petitjean et al. (1990).
According to our simulations, we speculate that the WD mixing fraction of this nova is greater than $40\%$.

Fig.\,4 also represents the WD mixing fractions measured by $\rm \sum CNO/Mg$ and $\rm \sum CNO/Si$.
We note that the measurement results of these two ratios only poorly agree with $\rm (H+He)/\sum CNO$ and $\rm (H+He)/Ne$
(for more discussions, see Sect. 4).
This figure shows that the WD mixing degree in classical novae spans a relatively wide range.
\begin{figure}
        \centering\includegraphics[width=\columnwidth*5/5]{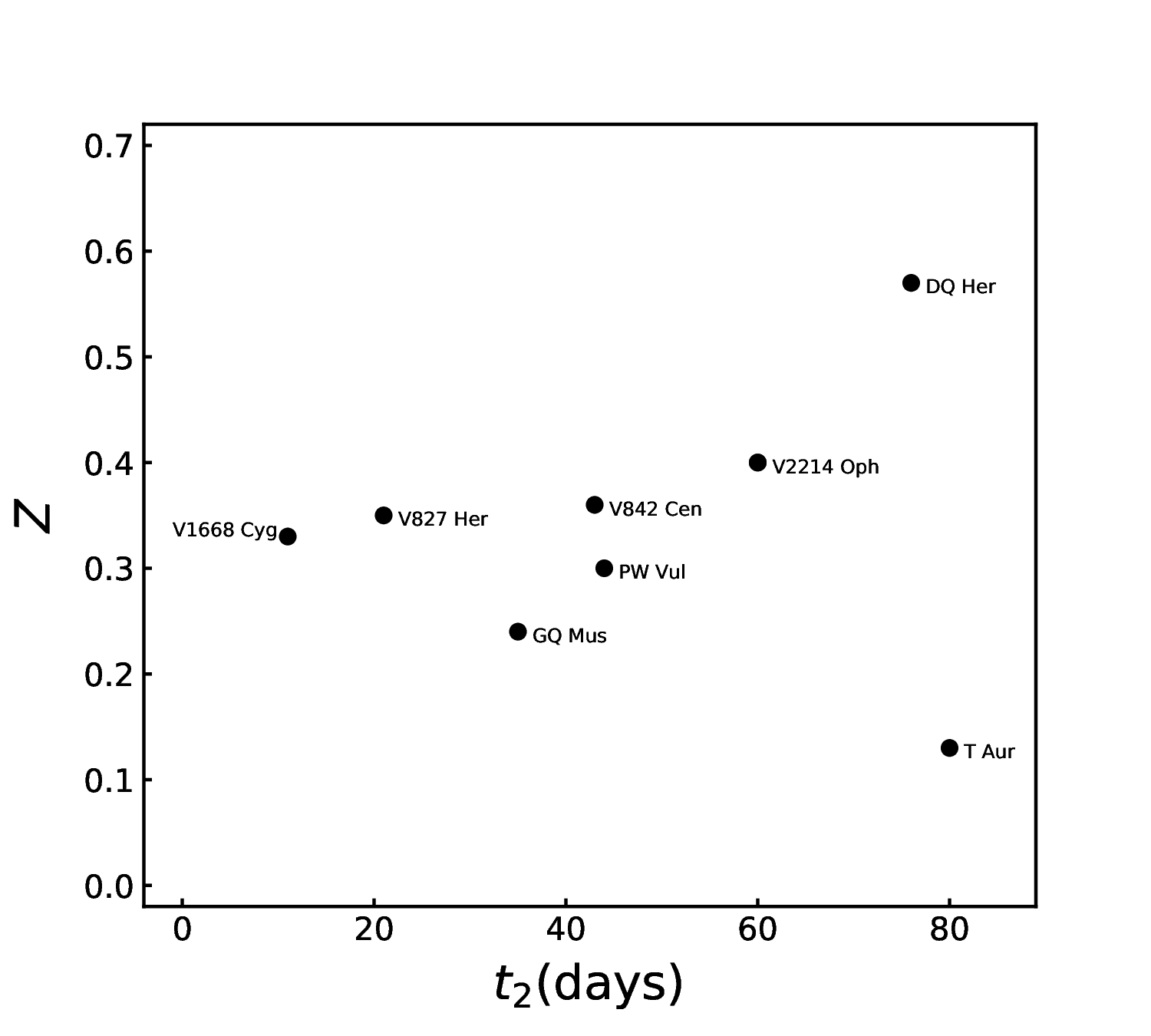}
        \caption{Relation between $t_{\rm 2}$ and the metallicity in nova ejecta ($Z$),
                in which these novae are listed in Gehrz et al. (1998).
                We only selected classical novae that show significant signs of the WD mixing ($Z \textgreater 0.1$). 
                $t_{\rm 2}$ is the time of decline by two magnitudes from peak brightness,
                which can be directly derived from the observations.}
        \label{fig:t2z}
\end{figure}
\subsection{Relation between $t_{\rm 2}$ and the metallicity in CO novae}
In order to explore the physical parameters that affect the WD mixing fraction,
we summarised the observed data of CO novae listed in Gehrz et al. (1998),
in which the selected novae show obvious signs of the WD mixing (i.e. the metallicity $Z \textgreater 0.1$).
Fig.\,5 shows the relation between $t_{\rm 2}$ and $Z$ in these classical novae,
where $t_{\rm 2}$ is the time of decline by two magnitudes from peak brightness.
We used $t_{\rm 2}$ instead of WD mass or accretion rate because $t_{\rm 2}$ can be obtained directly from the observations,
and the metallicity was chosen because it can reflect the WD mixing fraction in classical novae.

As shown in Fig.\,5, the longer $t_{\rm 2}$ time tends to correspond to higher metallicities in classical novae.
It is generally thought that a faster-declining nova that has a shorter $t_{\rm 2}$ implies a more massive WD in the nova system.\footnote{Other parameters
        (e.g. the accretion rate, the WD initial luminosity, and the composition of the accreted envelope)
        also affect $t_{\rm 2}$, but the WD mass affects the main trend most (e.g. Yaron et al. 2005; Hachisu \& Kato 2016; Hachisu et al. 2020).}
Thus, the trend illustrated in Fig.\,5 indicates that
a high metallicity (i.e. high WD mixing fraction) may exist in nova systems with less massive WDs,
except for the particular object nova T Aur, i.e. a low metallicity, but with $t_{\rm 2}$ close to that of DQ Her.

The WD mass would affect the mixing time, and this in turn affects the metallicity in classical novae.
It has been suggested that the temperature at the envelope base ($T_{\rm base}$)
is about $10^8$\,K when the convection has already extended throughout the whole envelope
(e.g. Jos{\'e} et al. 2016).
In multidimensional simulations,
Casanova et al. (2011) mapped a 1D nova model on a
2D or 3D Cartesian grid when $T_{\rm base}$ reached $10^8$\,K by considering multidimensional mixing
(see also Casanova et al. 2018).
Jos{\'e} \& Hernanz (1998) studied the effect of the WD mass on the time
required from $10^8$\,K to
the peak temperature ($t_{\rm max}$).
They found that $t_{\rm max}$ decreases with the WD mass,
which means that a shorter mixing time exist in nova systems with massive WDs,
resulting in a low WD mixing degree.

\section{Discussion}\label{disc}
As shown in Fig.\,4,
the WD mixing degree measured by different mixing meters may be inconsistent.
It has been suggested that the elements C and O are the main components in the underlying CO WDs
based on studies of stellar evolution and asteroseismology
(e.g. Umeda et al. 1999; Giammichele et al. 2018).
Consequently,
the WD material mixed with the accreted envelope is mainly composed of C and O,
indicating that $\rm (H+He)/\sum CNO$ is the most suitable elemental abundance ratio for estimating the WD mixing level.
On the other hand,
the abundances of Mg and Si in CO novae are relatively low,
hence they are not good candidates for the WD mixing meters.

According to stellar evolution theories, $^{22}\rm Ne$ is produced by $^{14}\rm N(\alpha,\gamma)^{18}\rm F$ $(e^+v)$ $^{18}\rm O(\alpha,\gamma)^{22}\rm Ne$ when stars evolve to the He-burning phase, which means that $^{22}\rm Ne$ becomes the dominant neon isotope in CO WDs. We set the mass fraction of $^{22}\rm Ne$ in the WD material to 0.01.
However, the $^{22}\rm Ne$ abundance in CO WDs may be different from our assumptions due to the different metallicity environments (Umeda et al. 1999),
which will affect the WD mixing degrees measured by $\rm (H+He)/Ne$.
For example,
V723 Cas has an Ne abundance as high as the abundances of ONe novae,
but it was identified as a CO nova (see Iijima 2006).
This high abundance of Ne is due to the high concentration of $^{22}\rm Ne$ in the underlying CO WD
(Livio \& Truran 1994; Hachisu \& Kato 2016), 
and it means that the WD mixing degree measured by $\rm (H+He)/Ne$ is much higher than $40\%$ (see Fig.\,4e).

\begin{figure}
        \centering\includegraphics[width=\columnwidth*5/5]{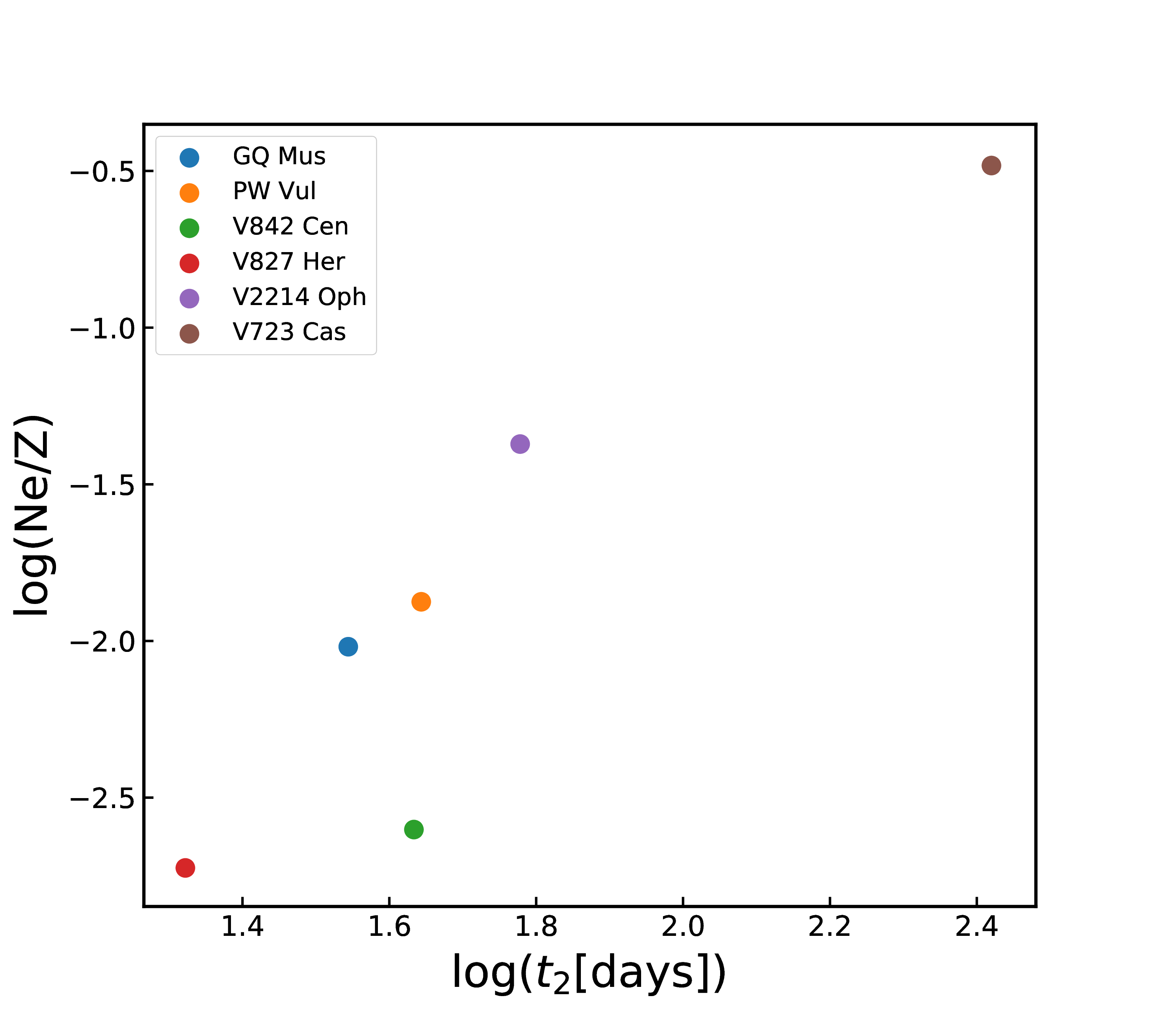}
        \caption{Relation between $t_{\rm 2}$ and the value of $\rm Ne/Z$ obtained from some classical novae,
                in which these novae show obvious signs of WD mixing ($Z \textgreater 0.1$).}
        \label{fig:lnez}
\end{figure}
Table 2 shows that the ratio of Ne abundance to the metallicity ($\rm Ne/Z$) in nova ejecta weakly dependends on the WD mass and the WD mixing degree,
and its value is consistent with the Ne abundance set in the CO WD model (i.e. $X(^{22}\rm Ne)$ = 0.01).
Thus, we expect that the Ne abundance in WDs ($X(^{22}\rm Ne)$) can be represented by the value of $\rm Ne/Z$ in nova ejecta.
Fig.\,6 shows the relation between $t_{\rm 2}$ and $\rm Ne/Z$ obtained from six classical novae,
where these novae show obvious signs of WD mixing ($Z \textgreater 0.1$).
Some other novae (e.g. V1668 Cyg, DQ Her and T Aur) are not presented here
because no information about the Ne abundance is available in the references
(e.g. Gallagher et al. 1980; Petitjean et al. 1990; Andre{\"a} et al. 1994).
As shown in Fig.\,6, 
lower $\rm Ne/Z$ (i.e. $X(^{22}\rm Ne)$) is preferentially detected in the faster nova systems (i.e. shorter $t_{\rm 2}$), corresponding to the massive WDs.
Previous studies suggested that as the stellar population metallicity increases,
the $^{22}\rm Ne$ abundance increases while the WD mass decreases
(see, e.g. Umeda et al. 1999).
Thus,
we speculate that these novae shown in Fig.\,6 may have different stellar population metallicities.

By assuming the mixing process driven by Kelvin-Helmholtz instabilities in the multidimensional simulations,
Casanova et al. (2018) suggested that the mixing fraction in nova systems increases with the WD mass
due to the higher base density and peak temperatures.
However, in their simulations,
the metallicities in novae with less massive WDs are lower than the metallicity of the observed values,
for instance $Z = 0.0268$ for $0.8\,M_\odot$, and $Z = 0.0772$ for $1\,M_\odot$ in CO novae.
We note that the nova models calculated in their work
involve a limited number of combinations of WD masses and accretion rates.
Thus, the influence of WD mass and other parameters such as the mass-accretion rate
on the WD mixing deserves further analysis.

The observed He enrichments in classical novae imply that helium can be accumulated on the WD surface.
We speculate that the WD mass could affect the thickness of He shell,
in which the heavier He shell is expected to accumulate on the surface of low-mass WDs
owing to the weaker outbursts caused by the relatively low degeneracy of the H shell.
Chen et al. (2019) studied the characteristics of nova outburst
by considering a wide range of WD masses and mass-accretion rates.
They found that low-mass WDs can lead to relatively high accreted masses
and mass-accumulative efficiencies for a given mass-accretion rate.
In addition, Glasner et al. (2012) showed that the He mixing level in nova systems is about $20\%$,
indicating that the mixed He shell mass is $\sim20\%$ of the critical mass.

\section{Summary}
By employing the stellar evolution code MESA,
we investigated the elemental abundance ratios during nova outbursts. These ratios can be used to estimate the WD mixing fraction in CO novae.
They show a weak dependence on the WD mass and the He mixing,
but they are sensitive to the WD mixing fraction.
We identified the elemental abundance ratios
$\rm (H+He)/\sum CNO$, $\rm (H+He)/Ne$, $\rm \sum CNO/Mg,$ and $\rm \sum CNO/Si$
that can meet the conditions for determining the WD mixing degree,
where $\rm (H+He)/\sum CNO$ is the most useful mixing meter.
In addition, we estimated the WD mixing fraction in some representative classical novae.
Moreover, we found that a higher metallicity (i.e. higher WD mixing fraction)
is preferentially accompanied by a longer $t_{\rm 2}$ time
in classical novae.
Our work can be used to provide some constraints on the mixing theories in nova systems.
In order to better understand the mixing process,
more observed samples and theoretical simulations are needed.

\begin{acknowledgements}
We thank the anonymous referee for valuable comments
that help to improve the paper.
BW is supported by the National Natural Science Foundation of China (No 11873085), the science research grants from the China Manned Space Project (Nos CMS-CSST-2021-A13/B07), the Western Light Youth Project of CAS, and the Yunnan Fundamental Research Projects (Nos 2019FJ001 and 202001AS070029).
CW is supported by the National Natural Science Foundation of China (No. 12003013).
The authors gratefully acknowledge the ``PHOENIX Supercomputing Platform'' jointly operated by the Binary Population Synthesis Group and the Stellar Astrophysics Group at Yunnan Observatories, Chinese Academy of Sciences.
\end{acknowledgements}

\end{document}